\documentclass[conference,10pt]{IEEEtran}
\usepackage[utf8]{inputenc}
\usepackage{color,xcolor}
\usepackage{epsfig}
\usepackage{amsmath,amsfonts,amsthm,amssymb}
\usepackage{graphics,graphicx,pgfplots,tikz,subfigure}
\usepackage{float}
\usepackage{algorithm,algorithmic}
\usepackage{svg}
\usepackage{epstopdf}
\usepackage{soul,csquotes}%,ulem}

\newcommand{\FRR}{F_{\text{RR}}}
\newcommand{\FPQ}{F_{\text{PQ}}}
\newcommand{\FOL}{F_{\text{OL}}}
\newcommand{\FVI}{F_{\text{VI}}}
\newcommand{\FDQA}{F_{\text{DQ}^{(1)}}}
\newcommand{\FDQB}{F_{\text{DQ}^{(2)}}}
\newcommand{\Ftarget}{F_{\text{trgt}}}

\newcommand{\figref}[1]{Fig. \ref{#1}}

\begin{document}
\title{An MDP approach for radio resource allocation\\in urban Future Railway Mobile Communication System (FRMCS) scenarios}
	\author{
		\IEEEauthorblockN{
			V. Corlay and J.-C. Sibel
		}
		\IEEEauthorblockA{
			Mitsubishi Electric R\&D Centre Europe\\
			Rennes, France\\
			Email: \{v.corlay, j.sibel\}@fr.merce.mee.com
		}
	}
	\maketitle
	
%====================================================================================================================================================
\begin{abstract}
  In the context of railway systems, the application performance can be very critical and the radio conditions not advantageous. Hence, the communication problem parameters include both a survival time stemming from the application layer and a channel error probability stemming from the PHY layer. 
This paper proposes to consider the framework of Markov Decision Process (MDP) to design a strategy for scheduling radio resources based on both application and PHY layer parameters.
The MDP approach enables to obtain the optimal strategy via the value iteration algorithm. The performance of this algorithm can thus serve as a benchmark to assess lower complexity schedulers. 
We show numerical evaluations where we compare the value iteration algorithm with other schedulers, including one based on deep Q learning.
%Considered as an optimal strategy regarding a long-term average performance, the MDP is the reference that any low complexity scheduler should compare with. 
\end{abstract}


%====================================================================================================================================================
\section{Introduction}
%====================================================================================================================================================
	  On the one hand, the automated train control is a crucial railway service use case and induces a change of communication paradigm compared to the current railway system: It might require at the same time a large throughput, a very high reliability, and a sufficient availability.  On the other hand, the success of the 3GPP 5G NR standard for communication systems makes the underlying technology relevant for specific scenarios such as that of the railway systems. As a result, the Future Railway Mobile Communication Systems (FRMCS) propose mechanisms to take advantage of 5G-related aspects to offer specific railway services such as the automated train control \cite{3GPP22.289}.
	
	Within this context, the scheduling of radio resources plays an important role to efficiently share the said resources between several users. The Round-Robin \cite{ArpaciDusseau18-Book} and the Priority-Queue \cite{Liebeherr1999} are well-known schedulers whose computational complexity is very low but whose performance is not optimal with respect to application-level metrics. As a matter of fact, they do not take into account all the parameters impacting the application performance. For the current purpose, the objective is to consider both application-level parameters and lower layer parameters, such as the radio conditions, to adapt the scheduling strategy.  %This scheduler scheduler performs allocation action taking into account the application performance metric that is the real target of the communication system. %The evaluations results are encouraging therefore we propose to continue in this direction.

	This approach is in line with the Release 19 of 3GPP \cite{3GPP22.261} which specifies the service requirements for the 5G system. In this latter reference,
	%we can indeed find the notion of the aforementioned \textit{survival time} being defined as \enquote{the time that an application consuming a communication service may continue without an anticipated message}. 
	it is explained that \enquote{the communication service is considered unavailable if it does not meet the pertinent Quality of Service (QoS) requirements. For example, the communication service is unavailable if a message is not correctly received within a specified time, which is the sum of maximum allowed end-to-end latency and survival time}. %Consequently, the proposal of \cite{Sibel2023} is clearly a good candidate. 

	Recently, we introduced a new paradigm \cite{Sibel2023} for the design of a radio resource scheduler in a multi-agent setting. It takes into account in the meantime an application layer parameter, the survival time, and a PHY layer parameter, the channel error probability. To enhance the scheduler of \cite{Sibel2023}, a low-complexity heuristic to approximately solve the scheduling optimization problem, we formalize in this paper the scheduling problem as a Markov Decision Process (MDP) \cite{Puterman1994,Sutton2020}. One advantage is that MDP provide an optimal solution for the optimization problem. Another advantage is that MDP have been widely experienced and give also access to many sub-optimal existing algorithms. More specifically, we shall consider the value iteration algorithm as well as the deep Q learning algorithm. The first algorithm optimally solves the scheduling problem but with a prohibitive complexity as the problem size grows. The second algorithms is sub-optimal but scales with the problem size, similarly to the heuristic proposed in \cite{Sibel2023}. Consequently, we compare these algorithms and discuss the performance complexity trade-off.
	
	This paper is organized as follows. Section \ref{sec:FRMCSContex} presents the scheduling problem in the scope of FRMCS, Section~\ref{sec:SystemDescription} describes the system, Section~\ref{sec:MDP} introduces MDP within the scheduling framework, and Section~\ref{sec:NumericalEvaluations} exhibits numerical results to challenge MDP with other schedulers.

%====================================================================================================================================================
\section{FRMCS context\label{sec:FRMCSContex}}
%====================================================================================================================================================
	Within the various scenarios under the umbrella of the train control \cite{3GPP22.289}, we focus on the remote driving for train shutting yards. This implies a remote driver driving the train in order to bring it back to the train station. The data provided to the said driver is mainly made of images and videos to allow the driver to be aware, in real-time, of the train surrounding environment. In other words, the data to be transmitted involves a large quantity of payloads. Moreover, as this environment is shared with other trains, moving or not, remote driving raises a safety problem. 

	Combining a high throughput with a high reliability and a high availability is one of the 5G proposals, e.g., for the V2X scenarios \cite{3GPP22.886} or the factory automation scenarios \cite{3GPP22.104}. The application requirements provided by the 3GPP specifications embody the performance target for the access layer design through the QoS. The access layer is understood in this paper to comprise the PHY and MAC layers. For example, the QoS comprises a guaranteed bit-rate, a latency, etc. However, the metrics that are used as inputs and outputs for the mechanisms of the access layer are low-level metrics, e.g., the channel error probability, the frame error rate, the channel busy/occupancy ratio. Even though these metrics are helpful to underline the behaviour of a single layer, namely the access layer, they do not well reflect the expected synergy with the application layer. 
	
	From another perspective, some companies provide in \cite{ETSI103.554-2} results of radio performance for NR railway systems considering a system-level framework. Within this framework, the scheduler plays an important role to efficiently share the radio resources between the various users. The said companies only consider the Round-Robin scheduler \cite{ArpaciDusseau18-Book}, a scheduler that assigns equal amount of resources to each user, regardless of the channel quality or performance requirements for the applications. Among other consequences, a user with a high channel quality achieves a much higher throughput than what it needs while, at the same time, a user with a low channel quality cannot achieve the required throughput. 
%This kind of strategy is suboptimal in spite of the specification tools that the QoS proposes to adapt the scheduling strategy. 
Furthermore, a proportional-fair strategy would bring only limited benefits as it does not consider the application parameters. Consequently, following the scope in \cite{Sibel2023}, we orient the current paper on the scheduling problem taking into account application aspects as well as access layer aspects.

%====================================================================================================================================================
\section{Description of the system\label{sec:SystemDescription}}
%====================================================================================================================================================
	This section presents the scheduling problem by defining the application traffic model, the radio resources used by the scheduler and the application behaviour, similarly to \cite[Sec. II.A]{Sibel2023}. The difference with \cite{Sibel2023} is the inclusion of the payload size, and the division of the payload in several packets.

	%====================================================================================================================================================
	\subsection{Traffic model}
	%====================================================================================================================================================
		We consider a discrete-time system divided in time slots whose length is $dt$, e.g., $dt=1$ msec. Let $N$ be the number of agents in the system. Two main traffic models for the application layer are commonly proposed in the literature \cite{Ameigeiras2012}: 
		\begin{itemize}
			\item \textit{Full buffer traffic model}: The buffers of an agent have always an unlimited amount of data to transmit. 
			\item \textit{Finite buffer traffic  model}: An agent is assigned a finite payload to transmit when it arrives. The agent leaves the system when the payload reception is completed.
		\end{itemize}
		In this study, we consider a \textit{full-finite buffer traffic model}. An agent $A_k$ is assigned a finite payload of size $P_k$ to transmit. As soon as the payload has been fully received, the application buffer of $A_k$ is immediately refilled with a payload of the same size. In this model, all the agents are always active as they have always a payload in their buffer.
	
	%====================================================================================================================================================
	\subsection{Scheduling resource \& Payload aspect}
	%====================================================================================================================================================
		We consider a single radio resource per time slot $t$. All agents simultaneously compete for the resource at $t$ and only one agent finally obtains it. The channel error probability $p_k$ is the probability for the agent $A_k$ that a transmission at any time $t$ fails because of the channel. It is assumed to be constant over time. 
		
		We assume that only a packet of size $\Gamma_k$ can be transmitted in one time slot for $A_k$. The payload is thus necessarily split into $C_k$ packets at the scheduler level, i.e., $P_k = C_k\Gamma_k$. The payload is transmitted once the associated packets have all been successfully transmitted. 
	
	%====================================================================================================================================================
	\subsection{Application behaviour and performance metric}
	%====================================================================================================================================================
		The application is monitored according to three events for any agent. We assume that the agents use the same application, i.e., they have the same survival time $\tau$:
		\begin{itemize}
			\item The event \textit{survival time failure} for an agent (E1): \enquote{No payload is successfully transmitted during the last $\tau$ time slots} where $\tau$ is a strictly positive integer. For an agent $A_k$ at time $t$, we accordingly introduce $\tau_k(t)$ as the remaining time before (E1). This means that $0 \leq \tau_k(t) \leq \tau$ and $\tau_k(t)$ is a decreasing function of $t$. 
			\item The micro-event \textit{successful transmission of a packet at time $t$} (e0).
			\item The macro-event \textit{successful transmission of a payload at time $t$} (E0): This event is the result of $C_k$ events (e0) required to to transfer a whole message.
		\end{itemize}
		With both the events (E0) and (E1) for any agent $A_k$ at time $t$, $\tau_k(t)$ is immediately set to its maximum value $\tau$. 
		%The traffic model, the scheduling resource and the events (e0) and (E0) are displayed in \figref{fig:AgentRepresentation}.
		%\begin{figure*}[!h]
			%\centering
			%\includegraphics[width=1.75\columnwidth]{Figures/TrafficModel.pdf}
			%\caption{Representation of the traffic full-finite buffer traffic model for agent $A_k$, the scheduling resource and the events (e0) and (E0). When a message $m^{(n)}$ arrives in the buffer, it is split into packets $p_1^{(n)},p_2^{(n)},\dots$. The packets play in the scheduling process one-by-one such that the packet $p_{i+1}^{(n)}$ plays as soon as the packet $p_i^{(n)}$ has been successfully transmitted. Once all the packets have been successfully transmitted through the channel, i.e., after $C_k$ occurrences of (e0), then comes the event (E0) and the application delivers a new message.}
			%\label{fig:AgentRepresentation}
		%\end{figure*}
			
For a single agent $A_k$, at any time $t$, the quantity $V_k(t)$ is the number of failures (E1) met by $A_k$ until time $t$. The performance metric is chosen as the failure rate $F(t)$:
		\begin{align}
			F(t) = \frac{\sum_{k=1}^N V_k(t)}{t}.
			\label{eq:OverallProbabilityOfFailure}
		\end{align}
		Given that several agents might fail at a single time slot, the value of the failure rate can be greater than one. This is to expect for very small values of the survival time $\tau$. 
%For a sufficiently long time duration, we consider that $F(t)$ has converged to a steady value $F$, i.e., that the total amount of failures $\sum_{k=1}^N V_k(t)$ linearly increases with $t$.
%====================================================================================================================================================
\section{Framework of the Markov Decision Process\label{sec:MDP}}
%====================================================================================================================================================
	We consider the infinite-horizon MDP with discounted sum-reward criterion, as presented in \cite[Chap. 6]{Puterman1994}, to model the scheduling problem.
	
	%====================================================================================================================================================
	\subsection{Model}
	%====================================================================================================================================================
		As for a standard MDP, we use four variables to model the scheduling problem:
		\begin{itemize}
			\item \textit{State} $\textbf{S}_t \in \mathbb{N}^{N \times 2}$: The element $\textbf{S}_t(k,1)$ is the number $\tau_k(t)$ of remaining time slots at time $t$ before meeting a failure for agent $A_k$. The element $\textbf{S}_t(k,2)$ is the number $c_k(t)$ of remaining packets at time $t$ before the application message is fully transmitted. As $0 \leq\tau_k(t)\leq\tau$ and $1\leq c_k(t)\leq C_k$, the state $\textbf{S}_t$ belongs to a state space $\mathcal{S}$ whose size is $|\mathcal{S}| = (\tau+1)^N \prod_{k=1}^N C_k$.
			\item \textit{Action} $a_t \in \{0,1,...,N\}$: The index of the agent who gets the resource at time $t$. The action $a_t$ belongs to an action space $\mathcal{A}$ whose size is the number of agent $N$.
			\item \textit{Short-term reward} $r_t \in \{ -N,...,-1,0\}$: Minus the number of events (E1) at time $t$. %The short-term reward orients the MDP according to some basic strategy. As mentioned in \cite{Sutton2020}, the short-term reward should be as simple and as instantaneous as possible, i.e., it should give a very basic outcome according to one single transition from one state to another one state. To avoid a bias, then, this reward must not reflect the underlying goal of the MDP being to find the best path in the state space explained in the paragraphs below.%On Figure~\ref{fig:state_example} no agent has a failure but there will one at the next time slot (agent 1).
			\item \textit{Transition probability}: If $a_t=k$, the packet for $A_k$ is well received with probability $1-p_k$. Hence, $a_t$ leads to only two states $\textbf{S}_{t+1}$ with the non-zero transition probability $p(\textbf{S}_{t+1}|\textbf{S}_t,a_t)$.% In Figure~\ref{fig:state_example}, agent 2 has a green square: there is no transmission error (the event having the probability $1-P_e$).
		\end{itemize}

		The MDP introduces the \textit{long-term reward} or \textit{gain} $G_t$ at time $t$ defined as: 
		\begin{align}
			G_t = \sum_{t'=t}^\infty \lambda^{t'-t} r_{t'},
		\end{align}
		where $\lambda<1$ is the \textit{discount factor}. If $\lambda$ is close to 1, this quantity is almost the same as the (unnormalized) failure rate~\eqref{eq:OverallProbabilityOfFailure}. The goal for the scheduling problem is thus to find a \textit{policy} $\pi$, being the time sequence of allocation decisions, that maximizes the expected gain $\mathbb{E}[G_t]$. %The expectation takes place as $r_{t'}$ is not deterministic given the transition probabilities. 

		The MDP then defines $v^\pi(\textbf{S})$ the \textit{value of the state} $\textbf{S}$ under the policy $\pi$ as the expected gain given that $\textbf{S}_t=\textbf{S}$:
		\begin{align}
		\label{equ_value_state}
			v^\pi(\textbf{S}) = \mathbb{E}[G_t|\textbf{S}_t=\textbf{S}].
		\end{align}
		Then, a policy $\pi^*$ is optimal if:
		\begin{align}
			v^{\pi^*}(\textbf{S}) \geq v^\pi(\textbf{S}), \ \forall \textbf{S} \in \mathcal{S} \text{ and } \forall \pi \neq \pi^*.
		\end{align}
		Under the optimal policy, the value of the state $\textbf{S}$ can be expressed via the Bellman's equation as:
		\begin{align}
			v^{\pi^*}(\textbf{S})  = \max_{a \in \mathcal{A}} Q(\textbf{S},a),
		\end{align}
		where $Q(\textbf{S},a)$ is the \textit{state-action value} computed as:
		\begin{align}
			Q(\textbf{S},a) = \sum_{\textbf{S}' \in \mathcal{S}} p(\textbf{S}'|\textbf{S},a) \Bigl(r(\textbf{S}',\textbf{S}) + \lambda  v^{\pi^*}(\textbf{S}')\Bigr),
		\end{align}
		where $r(\textbf{S}',\textbf{S})$ is the reward obtained when going from $\textbf{S}$ to $\textbf{S}'$. %$p(\textbf{S}'|\textbf{S},a)$ denotes the transition probability from $\textbf{S}$ to $\textbf{S}'$ given the action $a$ and 
		%\begin{align}
			%Q(\textbf{S},a) =  R(\textbf{S},a) + \lambda \sum_{\textbf{S}' \in \mathcal{S}} p(\textbf{S}'|\textbf{S},a) v^{\pi^*}(\textbf{S}'),
		%\end{align}
		%where $p(\textbf{S}'|\textbf{S},a)$ denotes the transition probability from $\textbf{S}$ to $\textbf{S}'$ given the action $a$. $R(\textbf{S},a)$ is the average reward of $\textbf{S}$ given the action $a$ such that:
		%\begin{align}
			%R(\textbf{S},a) = \mathbb{E}[r|\textbf{S},a] = \sum_{\textbf{S}' \in \mathcal{S}} p(\textbf{S}'|\textbf{S},a) r(\textbf{S}',\textbf{S})
		%\end{align}
		%where $r(\textbf{S}',\textbf{S})$ is the reward obtained when going from $\textbf{S}$ to $\textbf{S}'$. 
		Given an optimal policy $\pi^*$, the optimal action to take when in a state $\textbf{S}$ is obtained as:
		\begin{align}
			a^* = \arg \max_{a \in \mathcal{A}} Q(\textbf{S},a).
		\end{align}

	\subsection{Value iteration}
	%====================================================================================================================================================
%		The Bellman's equation implies an infinite recursive computation between all the states of $\mathcal{S}$ which is not practical. 
		One standard approach to obtain \eqref{equ_value_state} under $\pi^*$  is to operate the \textit{Value Iteration} algorithm (VI) that consists in computing the state values in an iterative manner. Let us define $v^{(i)}(\textbf{S})$ as the approximate of $v^{\pi^*}(\textbf{S})$ at iteration $i$. We accordingly define the approximate of $Q(\textbf{S},a)$ at iteration $i$ as:
		\begin{align}
			\label{equ_value_iteration}
			Q^{(i)}(\textbf{S},a) = \sum_{\textbf{S}' \in \mathcal{S}} p(\textbf{S}'|\textbf{S},a) \Bigl(r(\textbf{S}',\textbf{S}) + \lambda  v^{(i)}(\textbf{S}')\Bigr),
		\end{align}
		which leads to:
		\begin{align}
			\label{equ_value_iteration_2}
				%\begin{array}{l}
					v^{(i+1)}(\textbf{S}) = \max_{a \in \mathcal{A}} \ Q^{(i)}(\textbf{S},a),	
				%\end{array}
		\end{align}
		%\begin{align}
			%\label{equ_value_iteration}
			%v_{t+1}(s) = \max_{a \in \mathcal{A}} \left\{ R(s,a) + \lambda \sum_{s_j \in S} p(s_j|s,a) v_t(s_j) \right\},
		%\end{align}
		where $v^{(i)}(\textbf{S})$ converges to $v^{\pi^*}(\textbf{S})$ for all $\textbf{S}$ \cite{Puterman1994}. The optimal action is then chosen as $a^{(I)}(\textbf{S}) = \arg \max_{a \in \mathcal{A}} \ Q^{(I)}(\textbf{S},a)$, where $I$ denotes the index of the last iteration. 
		
		As already mentioned, the VI finds the optimal policy with respect to $\mathbb{E}[G_t]$. Hence, modelling the scheduling problem in the MDP framework enables to get the optimal performance in the cases where the VI is tractable. As a direct consequence, the VI can be used as a benchmark to assess the performance of less complex heuristics.
		%The vectorized version of the VI equation is:
		%\begin{align}
			%\label{equ_value_iteration_vec}
			%V_{t+1}= \max_{a \in \mathcal{A}} \left\{ (R + \lambda V_t)\textbf{P}_a \right\},
		%\end{align}
		%where:
		%\begin{itemize}
			%\item $V_t = [v_t(s_1), ..., v_t(s_N)]$ is the vector collecting all state values.
			%\item $R = [r(s_1), ..., r(s_N)]$ is the reward vector.
			%\item $\textbf{P}_a$ is the transition probability matrix for action $a$ where the $\textbf{P}_a(j,i) = p(s_j|s_i,a)$.
		%\end{itemize}

%The usual practical rule is to stop the loop on $i$ when for every state $\textbf{S}$ of $\mathcal{S}$, we observe a numerical convergence, i.e., $|v^{(i)}(\textbf{S})-v^{(i-1)}(\textbf{S})| < \epsilon$, where $\epsilon$ is an arbitrary small value. 

	%====================================================================================================================================================
	\subsection{Deep Q learning}
	%====================================================================================================================================================
		Even though the VI provides the optimal estimate of the states value, it suffers from significant computational complexity and memory consumption as it needs to cover all the states in $\mathcal{S}$ of size $|\mathcal{S}|=(\tau+1)^N \prod_{k=1}^N C_k$. A practical alternative to the VI is the temporal-difference learning \cite[Chap.~6]{Sutton2020} on which $Q$-learning and deep $Q$ learning rely. 
		
		Instead of looping over all the states to estimate $v^{\pi^*}(\textbf{S})$, these algorithms walk across a few states within $\mathcal{S}$. In the standard approach, the system is in state $\textbf{S}_t$ at $t$ and an action $a_t$ is done. One then gets a reward $r(\textbf{S},\textbf{S}')$ and the new state $\textbf{S}'$ of the system at time $t+1$. It is therefore a Monte Carlo approach. The model to estimate $Q^{\pi^*}(\textbf{S},a)$ is updated via an error signal $\Delta_t $, obtained from two estimates of $Q^{\pi^*}(\textbf{S},a)$:
		\begin{align}
			\label{equ_error_signal_standard}
			\Delta_t =\hat{Q}(\textbf{S},a) - \hat{Q}'(\textbf{S},a)
		\end{align}
		where:
		\begin{itemize}
			\item $\hat{Q}'(\textbf{S},a) = r(\textbf{S},\textbf{S}')+\lambda \max_{a'} \hat{Q}(\textbf{S}',a')$ is a first estimate of $Q^{\pi^*}(\textbf{S},a)$ based on the observed reward and the subsequent state $\textbf{S}'$ obtained when taking the action $a$ in state~$\textbf{S}$, and where $\hat{Q}(\textbf{S}',a')$ is obtained via the current model.
			\item $\hat{Q}(\textbf{S},a)$ is a second estimate of $Q^{\pi^*}(\textbf{S},a)$ obtained via the current model.
		\end{itemize}
		With deep Q learning the model is a neural network which is trained via the gradient of the error signal with respect to the parameters of the neural network \cite{Mnih2015}.

%====================================================================================================================================================
\section{Numerical evaluations\label{sec:NumericalEvaluations}}
%====================================================================================================================================================
	This section presents some performance results of the VI and deep Q learning algorithms, and compare them with other schedulers.% after having provided information regarding the implementation scenario.
	%====================================================================================================================================================
	\subsection{Challengers}
	%====================================================================================================================================================
		We compare the MDP-based algorithms with three other schedulers:
		\begin{itemize}
			\item \textit{Round-Robin} (RR). It consists in allocating the resource to the $N$ agents following a buffer of the agents index. The said buffer is a random permutation of $[0,\dots,N-1]$. After $N$ allocations, the RR replaces its buffer with a new random permutation of $[0,\dots,N-1]$. This randomization prevents an agent from being always out at each period of the RR. The RR is a low complexity algorithm.
			\item \textit{Priority-queue} (PQ). It consists in allocating the resource to the agent whose number of remaining time slots before meeting a failure is the lowest, i.e.,: $a_t = \text{arg min}_k \tau_k(t)$. This scheduler is very easy to implement with no concern neither on the computational complexity nor on the memory consumption. Nevertheless, it does not use the channel error probabilities $p_k$  and the number of remaining packets $c_k(t)$. The PQ is a low complexity algorithm.
			\item \textit{On-line} (OL) \cite{Sibel2023}. It uses a heuristic $f_r(t,k)$ to estimate a probability of survival time failure for every agent $A_k$. The agent to allocate is then selected to minimize a subsequent global probability of failure. %of the local short-term probability of survival time failure (\enquote{resilience violation} in the text) 
%It is an on-line scheduler whose objective is to minimize an overall probability of failure $F(t)$ as defined below in \eqref{eq:OverallProbabilityOfFailure}. 
In \cite{Sibel2023}, there is no notion of division of the payload in $C_k$ packets. Therefore, we need to slightly modify the heuristic $f_r(t,k)=p_k^{\tau_k(t)}$ with: 
%of the local short-term probability of survival time failure (\enquote{resilience violation} in the text) $f_r(t,k)=p_k^{\tau_k(t)}$ with:
			\begin{align}
				f_r(t,k)=p_k^{\frac{\tau_k(t)}{c_k(t)}}.
			\end{align}
			%In \cite{Sibel2023} are proposed two schedulers, we consider the exact solver \enquote{On-line sum}. 
			%We do not consider any fairness target therefore we set $\alpha=0$. 
			The OL is a moderately low complexity algorithm.
		\end{itemize}		
	\subsection{Scenario and assumptions}
	%====================================================================================================================================================
		We consider a small-sized scenario with $N=3$ agents and with $\tau$ spanning from $N$ to 12. We assume the channel error probabilities to be $p_1=10^{-3},p_2=10^{-2},p_3=10^{-1}$, i.e., such that $A_1$ has a very good channel, $A_3$ suffers from a difficult channel, and $A_2$ is in between. Several situations are studied regarding the number of packets $C_1,C_2,C_3$ per payload for the agents denoted by the set $\{C_1,C_2,C_3\}$: $\{1,1,2\}$, $\{1,1,3\}$, $\{2,2,2\}$ and $\{2,2,3\}$. Consequently, the size of the state space varies from 128 to 26364 elements.
In these configurations, the system has the greatest channel error probability on the agent with the greatest number of packets per payload. 
This is a way to stress the scheduler by setting difficult situations. 
The purpose of the evaluations is to observe the performance dependence on the values of $\{C_k\}_k$ given $\{p_k\}_k$. We let all the schedulers run for a time duration of $10^6$ time slots to be able to observe failure rates up to around $10^{-5}$. We recall that several failures might occur at a given time slot according to \eqref{eq:OverallProbabilityOfFailure}. Hence, the performance metric is $F=F(10^6)$ (see \eqref{eq:OverallProbabilityOfFailure}).

		The challengers are then compared taking the value of $F$ for each of them, namely $\FRR$ for the Round-Robin, $\FPQ$ for the Priority-Queue, $\FOL$ for the On-Line, $\FVI$ for the VI, $F_\text{DQ}$ for the deep Q learning. To simplify the wording here after, we denote by $F_X(\{a,b,c\})$ the value of the failure rate for the scheduler $X\in\{\text{RR},\text{PQ},\text{OL},\text{VI}\}$ with the configuration $C_1=a,C_2=b,C_3=c$. 
		%\begin{figure}[!h]
			%\centering
			%\blankFigure
			%\caption{Total amount of failures for the challengers against the time duration.}
			%\label{fig:LinearAmoutOfFailure}
		%\end{figure}		
	
	%====================================================================================================================================================
	\subsection{Details for Deep Q learning\label{sec:DQ}}
	%====================================================================================================================================================		
		We consider deep Q learning for the case $\{C_1,C_2,C_3\}=\{2,2,3\}$. We consider two neural networks whose input is the state $\textbf{S}_t$. The first network NN$^{(1)}$ is trained with the true situation $p_1=10^{-3},p_2=10^{-2},p_3=10^{-1}$ considering $\tau=10$. 
With such parameters, $A_1$ and $A_2$ encounter errors at a very low frequency, which may be an issue as deep Q learning is a Monte Carlo algorithm (analysis validated by the experiments, see the next section).
%As we will observe in the next paragraph, for such a survival time value, $F$ reaches very low values. Error events are so rare that NN$^{(1)}$ is not sufficiently trained to well manage them.  
To mitigate this issue, we train a second network NN$^{(2)}$ in a situation with $p_1=10^{-2},p_2=10^{-1},p_3=0.5$ such that $A_1$ and $A_2$ have more channel errors, while keeping the hierarchy $p_1<p_2<p_3$. In other words, we introduce a model mismatch for the training. Deep Q learning with NN$^{(1)}$ is called DQ$^{(1)}$ and DQ$^{(2)}$ with NN$^{(2)}$.

The neural network used in the simulation is a residual neural network of 22 layers comprising a total of 26200 parameters. Note however that a smaller neural network with only 3000 parameters yields similar results. Such a large network is considered such that its size is not the performance bottleneck.
		
	%====================================================================================================================================================
	\subsection{Simulation results}
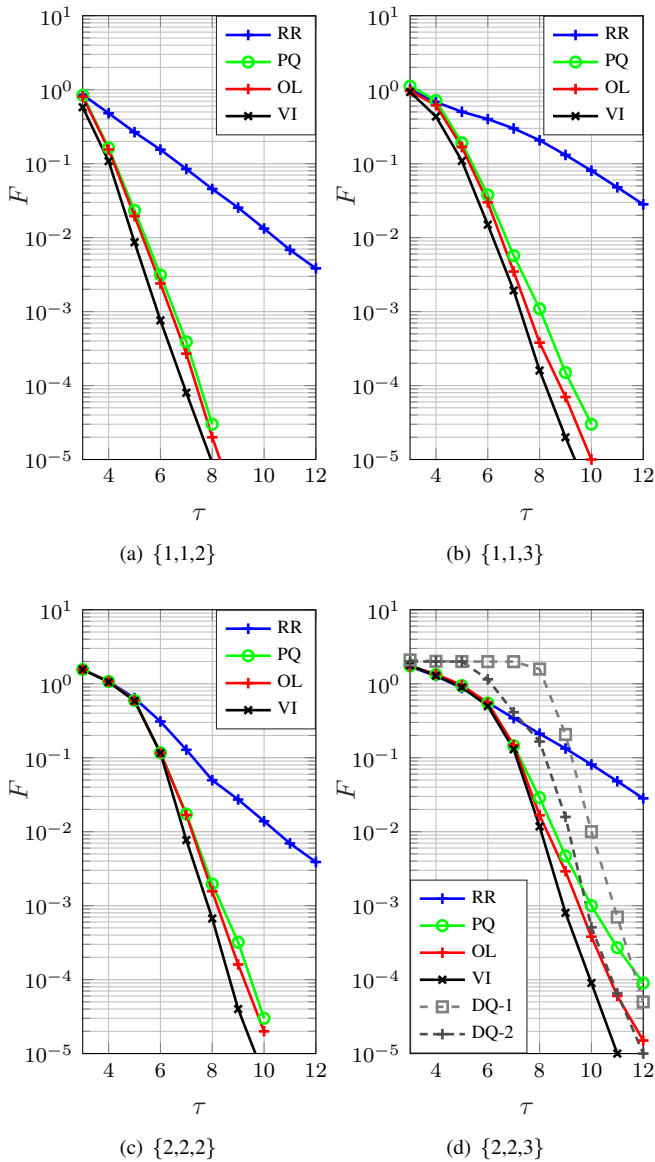
\begin{figure}[!h]
			\centering
			\subfigure[\{1,1,2\}]{% This file was created by matlab2tikz.
\definecolor{mycolor1}{rgb}{1.00000,1.00000,0.00000}%
\begin{tikzpicture}

\begin{axis}[%
	width=0.35\linewidth,
	height=0.25\textheight,
	font=\footnotesize,
	scale only axis,
	xmin=3,
	xmax=12,
	xlabel style={font=\color{white!15!black}},
	xlabel={$\tau$},
	ymode=log,
	ymin=1e-05,
	ymax=10,
	yminorticks=true,
	ylabel style={font=\color{white!15!black}, anchor=south west, at={(0.2,0.55)}},
	ylabel={$F$},
	axis background/.style={fill=white},
	title style={font=\bfseries},
	xmajorgrids,
	ymajorgrids,
	yminorgrids,
	legend style={at={(1,1)}, anchor=north east, legend cell align=left, align=left, draw=white!15!black, font=\scriptsize}
]
\addplot [color=blue, line width=1.0pt, mark size=2.0pt, mark=+, mark options={solid, blue}]
  table[row sep=crcr]{%
3	0.85498\\
4	0.48052\\
5	0.26636\\
6	0.1546\\
7	0.08462\\
8	0.04547\\
9	0.02541\\
10	0.01325\\
11	0.00682\\
12	0.00384\\
13	0.00202\\
14	0.00103\\
15	0.00058\\
16	0.00033\\
17	0.00022\\
18	0.00014\\
19	0.0001\\
20	6e-05\\
};
\addlegendentry{RR}

\addplot [color=green, line width=1.0pt, mark size=2.0pt, mark=o, mark options={solid, green}]
  table[row sep=crcr]{%
3	0.84086\\
4	0.16367\\
5	0.02339\\
6	0.00313\\
7	0.00039\\
8	3e-05\\%0\\
9	0\\
10	0\\
11	0\\
12	0\\
13	0\\
14	0\\
15	0\\
16	0\\
17	0\\
18	0\\
19	0\\
20	0\\
};
\addlegendentry{PQ}

\addplot [color=red, line width=1.0pt, mark size=2.0pt, mark=+, mark options={solid, red}]
  table[row sep=crcr]{%
3	0.80513\\
4	0.15542\\
5	0.0195\\%0.01095\\
6	0.0024\\
7	0.00027\\
8	2e-05\\
9	2e-6\\%2e-05\\
10	0\\
11	0\\
12	0\\
13	0\\
14	0\\
15	0\\
16	0\\
17	0\\
18	0\\
19	0\\
20	0\\
};
\addlegendentry{OL}

\addplot [color=black, line width=1.0pt, mark size=2.0pt, mark=x, mark options={solid, black}]
  table[row sep=crcr]{%
3	0.57865\\
4	0.10797\\
5	0.00866\\
6	0.00076\\
7	8e-05\\
8	9e-6\\%0\\
9	0\\
10	0\\
11	0\\
12	0\\
13	0\\
14	0\\
15	0\\
16	0\\
17	0\\
18	0\\
19	0\\
20	0\\
};
\addlegendentry{VI}

\end{axis}
\end{tikzpicture}%
}
			\hspace*{-1em}
			\subfigure[\{1,1,3\}]{% This file was created by matlab2tikz.
\definecolor{mycolor1}{rgb}{1.00000,1.00000,0.00000}%
\begin{tikzpicture}

\begin{axis}[%
	width=0.35\linewidth,
	height=0.25\textheight,
	font=\footnotesize,
	scale only axis,
	xmin=3,
	xmax=12,
	xlabel style={font=\color{white!15!black}},
	xlabel={$\tau$},
	ymode=log,
	ymin=1e-05,
	ymax=10,
	yminorticks=true,
	ylabel style={font=\color{white!15!black}, anchor=south west, at={(0.2,0.55)}},
	ylabel={$F$},
	axis background/.style={fill=white},
	title style={font=\bfseries},
	xmajorgrids,
	ymajorgrids,
	yminorgrids,
	legend style={at={(1,1)}, anchor=north east, legend cell align=left, align=left, draw=white!15!black, font=\scriptsize}
]
\addplot [color=blue, line width=1.0pt, mark size=2.0pt, mark=+, mark options={solid, blue}]
  table[row sep=crcr]{%
3	1.00515\\
4	0.68062\\
5	0.50286\\
6	0.40036\\
7	0.29917\\
8	0.20685\\
9	0.13112\\
10	0.0803\\
11	0.04787\\
12	0.02816\\
13	0.01562\\
14	0.0087\\
15	0.00511\\
16	0.00287\\
17	0.00155\\
18	0.00087\\
19	0.00045\\
20	0.00029\\
};
\addlegendentry{RR}

\addplot [color=green, line width=1.0pt, mark size=2.0pt, mark=o, mark options={solid, green}]
  table[row sep=crcr]{%
3	1.1171\\
4	0.71332\\
5	0.19142\\
6	0.03787\\
7	0.00572\\
8	0.00109\\
9	0.00015\\
10	3e-05\\
11	0\\
12	0\\
13	0\\
14	0\\
15	0\\
16	0\\
17	0\\
18	0\\
19	0\\
20	0\\
};
\addlegendentry{PQ}

\addplot [color=red, line width=1.0pt, mark size=2.0pt, mark=+, mark options={solid, red}]
  table[row sep=crcr]{%
3	0.95853\\
4	0.61649\\
5	0.16771\\
6	0.03\\%0.01432\\
7	0.00348\\
8	0.00038\\
9	7e-5\\%0.00011\\
10	1e-5\\%3e-05\\
11	0\\
12	0\\
13	0\\
14	0\\
15	0\\
16	0\\
17	0\\
18	0\\
19	0\\
20	0\\
};
\addlegendentry{OL}

\addplot [color=black, line width=1.0pt, mark size=2.0pt, mark=x, mark options={solid, black}]
  table[row sep=crcr]{%
3	0.9328\\
4	0.43373\\
5	0.10891\\
6	0.01496\\
7	0.00193\\
8	0.00016\\
9	2e-05\\
10	3e-6\\%0\\
11	0\\
12	0\\
13	0\\
14	0\\
15	0\\
16	0\\
17	0\\
18	0\\
19	0\\
20	0\\
};
\addlegendentry{VI}

\end{axis}
\end{tikzpicture}%
}\\
			\subfigure[\{2,2,2\}]{% This file was created by matlab2tikz.
\definecolor{mycolor1}{rgb}{1.00000,1.00000,0.00000}%
\begin{tikzpicture}

\begin{axis}[%
	width=0.35\linewidth,
	height=0.25\textheight,
	font=\footnotesize,
	scale only axis,
	xmin=3,
	xmax=12,
	xlabel style={font=\color{white!15!black}},
	xlabel={$\tau$},
	ymode=log,
	ymin=1e-05,
	ymax=10,
	yminorticks=true,
	ylabel style={font=\color{white!15!black}, anchor=south west, at={(0.2,0.55)}},
	ylabel={$F$},
	axis background/.style={fill=white},
	title style={font=\bfseries},
	xmajorgrids,
	ymajorgrids,
	yminorgrids,
	legend style={at={(1,1)}, anchor=north east, legend cell align=left, align=left, draw=white!15!black, font=\scriptsize}
]
\addplot [color=blue, line width=1.0pt, mark size=2.0pt, mark=+, mark options={solid, blue}]
  table[row sep=crcr]{%
3	1.55332\\
4	1.07091\\
5	0.63942\\
6	0.30866\\
7	0.12788\\
8	0.04966\\
9	0.02719\\
10	0.01384\\
11	0.00694\\
12	0.00389\\
13	0.00204\\
14	0.00103\\
15	0.00058\\
16	0.00033\\
17	0.00022\\
18	0.00014\\
19	0.0001\\
20	6e-05\\
};
\addlegendentry{RR}

\addplot [color=green, line width=1.0pt, mark size=2.0pt, mark=o, mark options={solid, green}]
  table[row sep=crcr]{%
3	1.55476\\
4	1.07297\\
5	0.59514\\
6	0.11627\\
7	0.01732\\
8	0.00196\\
9	0.00032\\
10	3e-05\\
11	0\\
12	0\\
13	0\\
14	0\\
15	0\\
16	0\\
17	0\\
18	0\\
19	0\\
20	0\\
};
\addlegendentry{PQ}

\addplot [color=red, line width=1.0pt, mark size=2.0pt, mark=+, mark options={solid, red}]
  table[row sep=crcr]{%
3	1.55606\\
4	1.07255\\
5	0.59501\\
6	0.11592\\
7	0.0168\\
8	0.00156\\
9	0.00016\\
10	2e-5\\%3e-05\\
11	0\\
12	0\\
13	0\\
14	0\\
15	0\\
16	0\\
17	0\\
18	0\\
19	0\\
20	0\\
};
\addlegendentry{OL}

\addplot [color=black, line width=1.0pt, mark size=2.0pt, mark=x, mark options={solid, black}]
  table[row sep=crcr]{%
3	1.55726\\
4	1.05611\\
5	0.58559\\
6	0.11546\\
7	0.0077\\
8	0.00067\\%0.00087\\
9	4e-5\\%2e-05\\
10	5e-6\\%0\\
11	0\\
12	0\\
13	0\\
14	0\\
15	0\\
16	0\\
17	0\\
18	0\\
19	0\\
20	0\\
};
\addlegendentry{VI}

\end{axis}
\end{tikzpicture}%
}
			\hspace*{-1em}
			\subfigure[\{2,2,3\}]{% This file was created by matlab2tikz.
\definecolor{mycolor1}{rgb}{0.5000,0.5000,0.5000}%
\definecolor{mycolor2}{rgb}{0.3000,0.3000,0.3000}%
\begin{tikzpicture}

\begin{axis}[%
	width=0.35\linewidth,
	height=0.25\textheight,
	font=\footnotesize,
	scale only axis,
	xmin=3,
	xmax=12,
	xlabel style={font=\color{white!15!black}},
	xlabel={$\tau$},
	ymode=log,
	ymin=1e-05,
	ymax=10,
	yminorticks=true,
	ylabel style={font=\color{white!15!black}, anchor=south west, at={(0.2,0.55)}},
	ylabel={$F$},
	axis background/.style={fill=white},
	title style={font=\bfseries},
	xmajorgrids,
	ymajorgrids,
	yminorgrids,
	legend style={at={(0,0)}, anchor=south west, legend cell align=left, align=left, draw=white!15!black, font=\scriptsize}
]
\addplot [color=blue, line width=1.0pt, mark size=2.0pt, mark=+, mark options={solid, blue}]
  table[row sep=crcr]{%
3	1.70349\\
4	1.27101\\
5	0.87592\\
6	0.55442\\
7	0.34243\\
8	0.21104\\
9	0.1329\\
10	0.08089\\
11	0.04799\\
12	0.02821\\
13	0.01564\\
14	0.0087\\
15	0.00511\\
16	0.00287\\
17	0.00155\\
18	0.00087\\
19	0.00045\\
20	0.00029\\
};
\addlegendentry{RR}

\addplot [color=green, line width=1.0pt, mark size=2.0pt, mark=o, mark options={solid, green}]
  table[row sep=crcr]{%
3	1.7349\\
4	1.3355\\
5	0.94508\\
6	0.54824\\
7	0.14525\\
8	0.02882\\
9	0.0047\\
10	0.001\\
11	0.00027\\
12	9e-05\\
13	0\\
14	0\\
15	0\\
16	0\\
17	0\\
18	0\\
19	0\\
20	0\\
};
\addlegendentry{PQ}

\addplot [color=red, line width=1.0pt, mark size=2.0pt, mark=+, mark options={solid, red}]
  table[row sep=crcr]{%
3	1.71051\\
4	1.35273\\
5	0.95393\\
6	0.551\\
7	0.14813\\
8	0.01664\\
9	0.00291\\
10	0.00038\\
11	6e-05\\
12	1.5e-05\\
13	0\\
14	0\\
15	0\\
16	0\\
17	0\\
18	0\\
19	0\\
20	0\\
};
\addlegendentry{OL}

\addplot [color=black, line width=1.0pt, mark size=2.0pt, mark=x, mark options={solid, black}]
  table[row sep=crcr]{%
3	1.77443\\
4	1.2867\\
5	0.89483\\
6	0.49806\\
7	0.1296\\
8	0.01176\\
9	0.0008\\
10	9e-05\\
11	1e-05\\
12	0\\
13	0\\
14	0\\
15	0\\
16	0\\
17	0\\
18	0\\
19	0\\
20	0\\
};
\addlegendentry{VI}

\addplot [color=mycolor1, line width=1.0pt, dashed, mark size=2.0pt, mark=square, mark options={solid, mycolor1}]
  table[row sep=crcr]{%
3	2.09654\\
4	2.01275\\
5	1.99513\\
6	1.9905\\
7	1.98811\\
8	1.57643\\
9	0.20614\\
10	0.01\\%0.00562\\
11	7e-4\\%0.00107\\
12	5e-5\\%1e-05\\
13	0\\
14	0\\
15	0\\
16	0\\
17	0\\
18	0\\
19	0\\
20	0\\
};
\addlegendentry{DQ-1}

\addplot [color=mycolor2, line width=1.0pt, densely dashed, mark size=2.0pt, mark=+, mark options={solid, mycolor2}]
  table[row sep=crcr]{%
3	1.99618\\
4	1.99417\\
5	1.99161\\
6	1.14616\\
7	0.41111\\
8	0.16524\\
9	0.0158\\
10	0.00051\\
11	6.5e-5\\%1e-05\\
12	1e-5\\%0\\
13	0\\
14	0\\
15	0\\
16	0\\
17	0\\
18	0\\
19	0\\
20	0\\
};
\addlegendentry{DQ-2}

\end{axis}

\end{tikzpicture}%
}
			\caption{Simulation results in several scenarios with $p_1=10^{-3},p_2=10^{-2},p_3=10^{-1}$.}
			\label{fig:Pe_1e-3_1e-2_1e-1}
		\end{figure}

	%====================================================================================================================================================		
		The results are displayed in \figref{fig:Pe_1e-3_1e-2_1e-1}. First of all, we observe that all schedulers have similar performance with large values of $F$ for low values of $\tau$. Unsurprisingly, a larger $\tau$  is required to reduce the failure rate. Moreover, the challenging zone for discriminating between the schedulers is then for middle or large values of $\tau$, e.g., for $\tau\geq 4$ with \{1,1,2\}, for $\tau\geq 5$ with \{1,1,3\}, for $\tau\geq 7$ with \{2,2,2\}, and for $\tau \geq 8$ with \{2,2,3\}. 
		
		Then, we observe that for such challenging zones, $\FRR$ is far beyond $\FPQ,\FOL$ and $\FVI$. This is expected as RR is the only scheduler which does not consider any agent parameter. Another common observation in every figure is that:
		\begin{equation}
			\FVI \leq \FOL \leq \FPQ \leq \FRR,
		\end{equation}
		This confirms that the VI is the reference scheduler. This also shows that the low complexity schedulers PQ and OL perform better than the naive RR. Consequently, the analysis will focus on the performance of OL and PQ in comparison with that of VI. 
		
		The configurations \{1,1,2\},\{2,2,2\} result in $\FOL\approx \FPQ$ whereas the other configurations \{1,1,3\},\{2,2,3\} result in $\FOL < \FPQ$ in a more obvious manner. As an example, at $\tau=10$, $\FPQ(\{1,1,3\}) = 3\FOL(\{1,1,3\})$ and at $\tau=12$, $\FPQ(\{2,2,3\}) = 4\FOL(\{2,2,3\})$. Also, the y-distance between $\FOL$ and $\FPQ$ increases when increasing $\tau$ in a faster way with $C_3=3$ than with $C_3=2$. For example, $\FPQ(\{2,2,2\})\approx 1.5\FOL(\{2,2,2\})$ at $\tau=10$ and $\FPQ(\{2,2,2\})\approx\FOL(\{2,2,2\})$ at $\tau=6$ while $\FPQ(\{2,2,3\})\approx 2.6\FOL(\{2,2,3\})$ at $\tau=10$ and $\FPQ(\{2,2,3\})\approx\FOL(\{2,2,3\})$ at $\tau=6$. This expansion effect is also observed between OL and VI, e.g., $\FOL(\{2,2,3\})\approx 4.2\FVI(\{2,2,3\})$ at $\tau=10$ and $\FOL(\{2,2,3\})\approx\FVI(\{2,2,3\})$ at $\tau=6$. As $\FOL\leq \FPQ$, though, OL can be said to be more robust than PQ when increasing the payload of the worse channel agent. In other words, enlarging the payload of the worse channel agent leads to select OL rather than PQ. 
		
		Alternatively, we can compare the schedulers with respect to the survival time they require to reach a target failure rate $\Ftarget$. As $\tau$ is an integer, we round $\tau$ to the nearest greater integer for comparison purpose. For example, when fixing $\Ftarget = 10^{-4}$, we see that for \{1,1,2\},\{1,1,3\},\{2,2,2\},\{2,2,3\}, respectively, VI fulfills the target with $\tau=7,8,9,10$, OL fulfills the target with $\tau=7,9,9,11$, and PQ fulfills the target with $\tau=8,9,10,12$. PQ is then always one time slot ahead from VI whereas OL may reach the same survival time. Consequently, OL provides nearly optimal performance with less constrains on the application compared with PQ.
		
		Finally, we focus on the curves showing the performance of deep Q learning with \{2,2,3\}. We observe that $\FDQA,\FDQB$ are both greater than $\FVI$, i.e., they do not reach the optimal policy. However, when $\tau\geq 11$, we see that $\FDQB\leq\FPQ$ and $\FDQB\leq\FOL$. Extrapolating the curves to $\tau\geq 12$, we expect DQ$^{(1)}$ and DQ$^{(2)}$ to perform even better than PQ and OL. In other words, when going further than the $\tau$ value used for training, DQ seems to well behave whereas running DQ for $\tau$ values lower than that of training, DQ is clearly suboptimal. We also observe that DQ$^{(2)}$, trained with the model mismatch, performs better than the standard training DQ$^{(1)}$. %We deduce that NN$^{(2)}$ is a more suited network and to enhance the performance, maybe an even more stressful training could be envisioned. 
This highlights that the rare channel errors is indeed a problem for deep Q learning, and having a model mismatch for the training improves the performance.
Nevertheless, this raises a new problem: How to chose the most adequate model parameters for the training? 
These preliminary performance results also indicate, though, that it might be possible to approach the VI performance provided good training parameters. Therefore, as DQ is less complex than VI and less memory consuming than VI, this opens the room for scaling the scheduling problem to a larger state space, e.g., with more agents, with greater values of $\tau$, and with greater values of $C_k$. 

\section{Conclusions}
%====================================================================================================================================================
	In this paper, we formalized the scheduling problem in the framework of MDP and we showed how to find the optimal scheduling strategy. This enables to assess the performance of candidate lower complexity schedulers. Indeed, the optimal scheduler suffers from complexity and storage issues. Among the lower complexity scheduler, the deep Q learning approach based on neural networks is investigated. We observed that training the neural network is not straightforward because of the scarcity of error events of some agents. Nevertheless, the performance can be improved by introducing a model mismatch for the training step. These preliminary results indicate that it may be possible to approach the optimal performance with this lower complexity scheduler. This may offer the possibility to scale the scheduler and therefore address larger systems.
%%========================================================================================
\bibliographystyle{IEEEtran}
\bibliography{IEEEabrv,biblio}
%%========================================================================================
%%====================================================================================================================================================
%\begin{thebibliography}{99}
	%\bibitem{3GPP22261}  XXX https://portal.3gpp.org/desktopmodules/Specifications/SpecificationDetails.aspx?specificationId=3107
	%\bibitem{Puterman1994} M. L. Puterman, ``Markov Decision processes: Discrete Stochastic Dynamic Programming," XXX, 1994.
	%\bibitem{Sibel2022} J.-C. Sibel, N. Gresset, and V. Corlay ``An application-oriented scheduler" XXX
	%\bibitem{Sutton2020} R. S. Sutton and A. G. Barto, ``Reinforcement Learning, An Introduction", 2nd ed.  XXX, 2020.
	%\bibitem{Ameigeiras2012} P. Ameigeiras, Y. Wang, J. Navarro-Ortiz, et al. ``Traffic models impact on OFDMA scheduling design". J Wireless Com Network 2012, 61 (2012).
%\end{thebibliography}
%%====================================================================================================================================================
\end{document}